# Lattice dynamic theory of compressed rare-gases crystals in the deformed atom model

V.N. Varyukhin[1], E.P.Troitskaya[1], V.V.Chabanenko[1], I.V. Zhikharev[1,2], E.E. Gorbenko[2], E.A. Pilipenko[1]

[1] Donetsk A.A. Galkin Physics and Technology Institute, National Academy of Science of Ukraine,

72 R. Luxemburg Str., 83114 Donetsk, Ukraine

[2] Lugansk Taras Shevchenko National University, 2 Oboronna Str., 91011 Lugansk, Ukraine

Lattice dynamics of rare-gas crystals is built on the base of adiabatic approximation when the deformation of electron shells of atoms of dipole and quadrupole types depending on nucleus shift and simultaneously arising Van-der-Vaals forces. The dipole forces are the most long-range ones. The obtained oscillation equations are studied in long-wave approximation. The role of three-body interaction and quadrupole deformation in the violation of Cauchy relation is discussed. Birch elastic moduli calculated for Xe and deviations from Cauchy relation are in good agreement with the experiment in a wide pressure range.

## Introduction

Rare-gas crystals (RGC) are relatively simple systems to study in comparison with other crystals because of their closed-shell electronic configuration and only one atom is in a unit cell. Special interest in RGC is connected with their properties under high pressures that make them applicable as a hydrostatic pressure medium in high-pressure diamond anvil cells (DAC) experiments[1].

Use of high-pressure Brilloin spectroscopy combined with DAC technology opened up new possibilities in intensive study of elastic properties of RGC within a wide pressure range [1,2,3,4]. The last paper of this series of ultra-precision measurements summarizes the results and contains a discussion, whether the existing theory adequately describes the experiment of deviation from Cauchy relation (CR). The authors of [5] noted that *ab initio* calculations of the density functional theory (DFT) [6] do not reproduce deviation from Cauchy relation δ even qualitatively. Their calculation of δ shows negative pressure dependence for all the series Ne, Ar, Kr, Xe with the pressure coefficient clearly depending on the atomic weight. The succession of Ne, Ar, Kr, Xe is observed in measurements of δ only under zero pressure. With pressure increase, as shown by experiments in [5], an individual pressure dependence of δ is observed. Thus $|\delta|$ of heavy crystals of Kr and Xe at p ≥ 10 GPa is less than $|\delta|$ of Ar. The calculations of the equation of state (EOS) and elastic moduli with empirical potentials within the frameworks of the many-body models demonstrate the inaccuracy less than 1% for each separate value and successfully reproduce the experimentally observed negative Cauchy relation, but $|\delta|$ of Xe is more of $|\delta|$ of Ar (see. [7] with references).

Such a discrepancy can be connected with the fact that the forces of one nature are formally substituted with the forces of another nature. Thus, *ab initio* calculations as well as empirical calculations of δ even using many-body interaction [7] principally differ from experimental values of δ for heavy RGCs.

This is an evidence that the description of strongly compressed substance requires a revision of the principal conceptions and approximations of the theory. It is clear that the theory must be *a priori* a microscopic and qualitative one (quantum mechanical) to be built on the first principles without fitting parameters. Because from the other hand, the system is exceptionally many-electron one, Hartree-Fock method can be used as a basic approach of the analysis. It is clearly formulated, accurate enough and it is not too complex to be realized using modern computers (see [8]).

The conception of interatomic interaction potential that plays such an important role in the investigation of crystal structure, lattice dynamics and thermodynamic properties is not a primary one as the conception of Coulomb charge interaction, for instance. It can be introduced and substantiated only in a definite approximation according to especially stipulated criteria.

The intrinsic electron energy together with the known Coulomb nucleus potential forms so called adiabatic potential of a nuclear system. Its rigorous description for the case of molecule was given in 1927 by Born and Oppenheimer [9] and in monograph of Born and Huang [10] that developed successive approach of adiabatic approximation applied to crystals.

While studying phenomena determined by crystal lattice dynamics and the processes of excitation and polarization of crystal atoms, we see their common basis, namely, the lowest energy level of the electron subsystem is an adiabatic potential of the motion of nuclei. Electron processes correspond to different levels of excitation of the same electron subsystem, which can be interpreted as quasiparticles able to interact one with another and with phonons, i.e. with elemental excitations of nuclear subsystem. Nevertheless, the majority of the theoretical studies ignores this primary relation and introduces electron-phonon (or exciton-phonon) interaction in a phenomenological way. Moreover, excited states of electron subsystem are necessary for the derivation of the adiabatic potential because the displacements of the nuclei from the equilibrium states adiabatically change the states of all the electrons within the crystal. This change can be naturally accounted for by the addition of a contribution of the excited states to the wave function of the basic state of electron subsystem. That was the way that realized adiabatic approximation in the pioneer works of K.B. Tolpygo [11,12,13,14]. Later, we applied the mentioned method to atomic cryocrystals (rare-gas crystals) [15] to studying short-range, noncentral and many-body forces in these crystals [16,17,18], in particular. First we considered only dipole atom deformation when the change of each atom state was characterized only by three parameters that were the components of dipole moment of its electron shell $\mathbf{P}_s^l$. That was the base of studying of a series of crystals [19,20,21]. To clarify a number of peculiarities of phonon spectrum of alkali-halogen crystals, it appeared necessary to include quadrupole atom deformation into consideration [22,23]. Such a consideration was extended to rare-gas crystals in [24]. Dealing with phonon spectrum, all these works can be classified as semi-empirical theories, because the parameters of the adiabatic potential were not calculated but can be found in different experiments [25,26] (classical version of K.B. Tolpygo's model accounting for atom polarizability and deformability a deformed atom model (DAM)). Nevertheless they have common quantum-mechanical basis: a method of realization of adiabatic approximation yielding a general form of adiabatic potential with the parameters expressed through definite matrix elements of electron subsystem Hamiltonian on atomic functions.

The present work is devoted to the study of interatomic interaction and dynamical lattice theory within the frameworks of the DAM based on the known wave function of the ground state of electron subsystem that is constructed, in turn, from the functions of the ground and the excited atom states. Here we require not the wave functions of the excited atom themselves but some integrals containing them. This method gives us an opportunity to *ab initio* calculate a number of crystal characteristics in a wide pressure range, to compare some calculated parameters with experimentally found ones at $p=0$. Besides, we can also get information about the interaction of elementary electron excitations with lattice oscillations (electron-phonon interaction in dipole and quadrupole approximation).

The first part is aimed at the generalization of the obtained results and formation of a uniform viewpoint on a wide circle of problems including lattice characteristics of RGC, in particular, their elastic properties. We shall shortly outline the way of derivation of the general form of the adiabatic potential from the primary principles. With all this going on, all the parameters are expressed in terms of matrix elements of Hamiltonian on the atomic wave function of the ground and the excited states of electron subsystem. The second part of the work shall contain the calculation of these parameters as functions of interatomic distance. The proposed theory is applied to investigate the high-pressure elastic properties and deviation from Cauchy relation of rare-gas crystal Xe.

# I. General theory. The deformed atom model (DAM). Adiabatic potential and oscillation equations of rare-gas crystals in a harmonic approximation.

## 1. Atomic shell deformation at lattice oscillations. Adiabatic potential.

As mentioned above, to obtain a proper adiabatic potential, it is necessary to find the wave function of electron subsystem $\Psi$ and its energy $U$ depending on the displacement of nuclei. It was supposed in [10] that it was done accurately. Here we shall try to introduce this dependence in some approximation. That is we define the ground state of the deformed crystal $\Psi_0$ in the form of a symmetrized product of the functions of separate atoms $\psi^l$.

$$\Psi_0 = A.c. \prod_l \psi^l \left( \mathbf{r}_1^l, ..., \mathbf{r}_N^l \right), \quad (1)$$

where $N$ is the number of electrons of each atom; A.c. means asymmetrization of the product. Every atomic function is supposed to be slightly different from $\psi_0^l$ function of an isolated atom and is presented as an expansion with respect to the lowest $k$ excited functions $\psi_i^l$

$$\psi^l = c_0^l \psi_0^l + \sum_{i=1}^k c_i^l \psi_i^l. \quad (2)$$

Here

$$\left| c_0^l \right|^2 + \sum_{i=1}^k \left| c_i^l \right|^2 = 1 \text{ and } \left| C_i^l \right| << 1 \quad (3)$$

First this method was suggested in [11,14] for alkali-halogen crystals and extended to the rare-gas crystals in [15].

Practical calculations with $\psi_i^l$ function require them to be orthogonal each other at varied $l, l'$. As $c_i^l$ are vanishing, we limit with the terms below $\left| c_i^l \right|^2$. This can be reached if we suggest the integrals on non-orthogonality to be small

$$S^{ll'} = \int \psi_0^l \psi_0^{l'} d\tau, \quad \sigma^{ll'} = \int \psi_i^l \psi_0^{l'} d\tau; \quad (4)$$

Thus we should first make all the functions $\psi_0^l$ orthogonal one to another and then to orthogonalize $\psi_i^l$ to already re-defined $\psi_0^{l'}$. With all this going on, $\psi_i^l$ is no more an eigenfunction of atomic Hamiltonian. Moreover, we shall not require the orthogonality of $\psi_i^l$ and $\psi_j^{l'}$ at $l \neq l'$, because it results in an error of higher infinitesimal order. We should note that the use of strong bond approximation for RGC will yield an error smaller than in the case of ion crystals where anion functions are less compact and strongly collides with cation functions and their excited states cannot be considered to be localized.

Function (1) accounting for (2) depicts the state of crystal atoms slightly distorted (in comparison with the isolated atom state) due to their integration into a lattice. Such distortion can be interpreted as small virtual non-correlated excitations of all the atoms.

Nevertheless this state cannot provide for the stability of a crystal built from neutral atoms because the repulsion of electron shells appears stronger of Coulomb and exchange interactions. The crystal is bound by Van der Waals forces arising as a result of virtual pair correlated excitations. In [15], we have described the corresponding states with the function

$$\Psi = c_0 \Psi_0 + \frac{1}{2} \sum_{ll'ij} c_{ij}^{ll'} \Psi_{ij}^{ll'}, \quad \left| c_{ij}^{ll'} \right| << 1, \quad (5)$$

where $\Psi_{ij}^{ll'}$ are function built from $\Psi_0$ (1) by the substitution of 2 factors of $\psi^l$ and $\psi^{l'}$ by the functions of excited states

$$\psi^l \to (\psi_i^l - c_i^l \psi_0^l); \quad \psi^{l'} \to (\psi_j^{l'} - c_j^{l'} \psi_0^{l'}). \quad (6)$$

The terms $-c_i^l \psi_0^l$ and $-c_i^{l'} \psi_0^{l'}$ provide the orthogonality of $\Psi_{ij}^{ll'}$ and $\Psi_0$. In practice, using the method of [15], we must not limit with the expansion (5), as the function can contain both pair excitation and triple, quadruple ones etc. So, if the corresponding coefficients $c_{ijk}^{ll'l''}$ and $c_{ijkm}^{ll'l''l'''}$ vanish as higher degrees of the ratio $M/\Delta E$, where $M$ are matrix elements of excitation and $\Delta E$ is the related energy of excitation, the number of terms will increase as $N(N-1)(N-2)/3!$ and $N(N-1)(N-2)(N-3)/4!$ (N is the number of atoms in the crystal). At the same time, it follows from the above reasoning that if we continue expansion (5), the main contribution to $\Psi$ will be made by the totality of the terms with large number of excitation $n \to \infty$, but only if the ratio $n/N \ll 1$ at $M/\Delta E \ll 1$. So we can make a conclusion that the account for atom interaction results in an amendment to energies in the form of a sum of Van der Waals interaction for each atom pair in the second order, a sum of Axilrode-Teller terms in the third order etc. Indeed, each term of Hamiltonian, containing more than two-particle interaction and applied to $\Psi$ function, produces virtual excitations in no more than two atoms. As the majority of atoms are not excited*, the terms

$$\left\langle 00 \left| \hat{H} \right|_{ij}^{ll'} \right\rangle \left\langle _{ij}^{ll'} \left| \hat{H} \right| 00 \right\rangle / (E_i - E_j - 2E_0)$$

make the main contribution to the amendment of the second order. The worded qualitative considerations can be supplemented with more rigorous argumentation based on many-body perturbation theory [27], as it was done in [28].

Adiabatic potential was obtained in the form [24]:

$$U \to \sum_l \left\{ \begin{array}{l} \dfrac{(\mathbf{P}^l)^2}{2\alpha} + \dfrac{1}{2} \sum_{\alpha\beta}^{9} \dfrac{(Q_{\alpha\beta}^l)^2}{2\beta_{44}} + \boldsymbol{\beta}^l \mathbf{P}^l + \dfrac{1}{2} \sum_{\alpha\beta} D_{\alpha\beta}^l Q_{\alpha\beta}^l \\ -\dfrac{1}{2} \sum_{l'} \left[ \dfrac{C}{|\mathbf{r}^{ll'}|^6} + \dfrac{C'}{|\mathbf{r}^{ll'}|^8} + \dfrac{C''}{|\mathbf{r}^{ll'}|^{10}} \right] \\ + \dfrac{1}{2} \sum_{l'} K(\mathbf{P}^l, Q_{\alpha\beta}^l, \mathbf{P}^{l'}, Q_{\alpha\beta}^{l'}) + \dfrac{1}{2} \sum_{l'}^{n.n.} U_{sr}(|\mathbf{r}^l - \mathbf{r}^{l'}|) \end{array} \right\} \quad (7)$$

Here $\sum_{\alpha\beta}^{(9)}$ means that we should look over all 9 combinations of α, β indexes (though only 5 of 9 components $Q_{\alpha\beta}^l$ are independent); $\sum_{l'}^{n.n.}$ is summation by the nearest neighbors;

$$\boldsymbol{\beta}^l = \frac{1}{\alpha} \sum_i \sum_{l'}^{n.n.} \frac{\left\langle 0 \left| \mathbf{P}^l \right| i \right\rangle \left\langle i0 \left| \hat{H}_{sr}^{ll'} \right| 00 \right\rangle + c.c.}{E_i - E_0};$$

$$D_{\alpha\beta}^l = \frac{1}{\beta_{44}} \sum_i \sum_{l'}^{n.n.} \frac{\left\langle 0 \left| \hat{Q}_{\alpha\beta}^l \right| i \right\rangle \left\langle i0 \left| \hat{H}_{sr}^{ll'} \right| 00 \right\rangle + c.c.}{E_i - E_0}. \quad (8)$$

Matrix elements of dipole and quadrupole moments are

$$\left\langle 0 \left| P^l \right| i \right\rangle = \int \psi_0^l \cdot P^l \psi_i^l d\tau,$$

$$\left\langle 0 \left| \hat{Q}_{\alpha\beta}^l \right| i \right\rangle = \int \psi_0^l \hat{Q}_{\alpha\beta}^l \psi_i^l d\tau, \quad (9)$$

here $K$ is Coulomb integral of interaction of all dipole and quadrupole moments written according to (9) as

$$P^l = \sum_i c_i^l \left\langle 0 \left| P^l \right| 0 \right\rangle + c.c.;$$

$$Q_{\alpha\beta}^l = \sum_i c_i^l \left\langle 0 \left| \hat{Q}_{\alpha\beta}^l \right| i \right\rangle + c.c.; \quad (10)$$

$$\frac{1}{2} \sum_{ll'} K(P^l, Q^l; P^{l'}, Q^{l'}) = \frac{1}{2} \sum_{ll'} \left\{ \begin{array}{l} \dfrac{\mathbf{P}^l \cdot \mathbf{P}^{l'}}{|\mathbf{r}^{ll'}|^3} - 3 \dfrac{(\mathbf{P}^l \cdot \mathbf{r}^{ll'})(\mathbf{P}^{l'} \cdot \mathbf{r}^{ll'})}{|\mathbf{r}^{ll'}|^5} \\ -2 \sum_{\alpha\beta} \dfrac{Q_{\alpha\beta}^l \cdot X_\alpha^{ll'} \cdot P_\beta^{l'}}{|\mathbf{r}^{ll'}|^5} + \\ +5 \sum_{\alpha\beta} \dfrac{Q_{\alpha\beta}^l \cdot X_\alpha^{ll'} \cdot X_\beta^{ll'} \cdot \mathbf{P}^{l'} \mathbf{r}^{ll'}}{|\mathbf{r}^{ll'}|^7} + \\ + \dfrac{25}{6} \sum_{\alpha\beta} \dfrac{Q_{\alpha\beta}^l \cdot Q_{\alpha\beta}^{l'}}{|\mathbf{r}^{ll'}|^5} - \\ - \dfrac{5}{3} \sum_{\alpha\beta\gamma} \dfrac{Q_{\alpha\beta}^l \cdot Q_{\beta\gamma}^{l'} X_\alpha^{ll'} \cdot X_\gamma^{ll'}}{|\mathbf{r}^{ll'}|^7} + \\ + \dfrac{35}{6} \sum_{\alpha\beta\gamma\delta} \dfrac{Q_{\alpha\beta}^l \cdot Q_{\gamma\delta}^{l'} X_\alpha^{ll'} \cdot X_\beta^{ll'} \cdot X_\gamma^{ll'} \cdot X_\delta^{ll'}}{|\mathbf{r}^{ll'}|^9} \end{array} \right\}.$$

(11)

where $\hat{H}_{sr}^{ll'}$ is the Hamiltonian of interaction of $l$ and $l'$ atoms, with separated long-range energy caused by Coulomb interaction of all dipoles and quadrupoles included to $K$ (11).

---

* Precisely, each atom is unexcited with overwhelming probability.

$$U_{sr}(|\mathbf{r}^l - \mathbf{r}^{l'}|) = \langle 00|\hat{H}_{sr}^{ll'}|00\rangle + \alpha(\boldsymbol{\beta}^l)^2 +$$
$$+ \sum_{\alpha\beta}^{(9)} \beta_{44}(D_{\alpha\beta}^l)^2 - 2(\sum_i \frac{1}{E_i - E_0} \sum_{l'} \langle 0i|\hat{H}_{sr}^{ll'}|00\rangle)^2. \quad (12)$$

The first term in (12) renders the interaction of two undistorted atoms and appears to be central if we do not account for the requirement of orthogonality of $\Psi_0^l$ and $\Psi_0^{l'}$. The rest of the terms will clearly contain three-body terms depending on coordinates, as seen from the designations, $\mathbf{r}^l, \mathbf{r}^{l'}, \mathbf{r}^{l''}$, where $l'$ and $l''$ are the nearest neighbors of $l$ site. So, the related interaction is eccentric.

## 2. Oscillation equations of RGC in harmonic approximation

The obtained adiabatic potential (7) is not absolute, but relative minimum of the average Hamiltonian. It corresponds to the least possible choice of electron function $\Psi$ from the view point of variation principle at some additional conditions when arbitrary values of dipole and quadrupole moments of all atoms $\mathbf{P}^l$, $Q_{\alpha\beta}^l$ (10) are fixed. Using (7), all coefficients $c_i^l$ are expressed by these elements (therefore, the form of $\Psi$ function, too). Thus we can consider them to be the only variation parameters left indefinite. To find them, it is necessary to minimize the expression of $U$ (7) with respect to $\mathbf{P}^l$, $Q_{\alpha\beta}^l$, and the following equations are obtained:

$$\frac{\partial U}{\partial P_\alpha^l} = 0, \qquad \frac{\partial U}{\partial Q_{\alpha\beta}^l} = 0. \quad (13)$$

The second group of equations (13) implies that only three of six types of mixed moments $Q_{\alpha\beta}^l$ are independent, $Q_{\alpha\beta}^l = Q_{\beta\alpha}^l$, and only two of three diagonal moments $Q_{\alpha\alpha}^l$ are independent, because $\sum_\alpha Q_{\alpha\alpha}^l = 0$. In practice, it is convenient to differentiate $U$ independently with respect to each $Q_{11}^l, Q_{22}^l, Q_{33}^l$, accounting for this additional condition. After elimination of $P_\alpha^l, Q_{\alpha\beta}^l$ from (13) and their substitution into $U$, we get the true adiabatic potential. The equation of motion with this potential is written as

$$m\ddot{u}_\alpha^l = -\frac{\partial U}{\partial u_\alpha^l}. \quad (14)$$

Because of (13), accounting only for explicit displacement dependence of $U$ is sufficient while differentiating $U$ with respect to $u_\alpha^l$. Hence it is convenient to consider equations (13), (14) together, suggesting $\mathbf{P}^l, Q_{\alpha\beta}^l$ to be additional dynamical parameters with the corresponding zero masses, though†.

Here we shall restrict the consideration to harmonic approximation when it becomes possible to estimate all parameters of the theory and to get intrinsic frequencies and lattice oscillation magnitudes. Addressing to (7), we see, that two first terms of the adiabatic potential are quadratic in $\mathbf{P}^l, Q_{\alpha\beta}^l$, as well as Coulomb interaction (11). Two next terms are linear in $\mathbf{P}^l, Q_{\alpha\beta}^l$. That is why their coefficients $\boldsymbol{\beta}^l, D_{\alpha\beta}^l$ should be presented in linear approximation with respect to displacement vectors $\mathbf{u}^l$. Finally, the value of $U_{sr}(\mathbf{r}^l - \mathbf{r}^{l'})$ and Van der Waals energy should be written in the form of quadratic expansion in terms of $\mathbf{u}^l$, $\mathbf{u}^{l'}$. It is seen from an expression for $\boldsymbol{\beta}^l$ (8), that it is a sum of $\boldsymbol{\beta}^{ll'}$ items over $l'$ with every item depending only from coordinate difference $(\mathbf{r}^l - \mathbf{r}^{l'})$. As it is a vector, and the only preferential direction is the vector of $(\mathbf{r}^l - \mathbf{r}^{l'})$, it is obvious that $\boldsymbol{\beta}^{ll'} = \frac{(\mathbf{r}^l - \mathbf{r}^{l'})}{|\mathbf{r}^l - \mathbf{r}^{l'}|} \cdot \left|\boldsymbol{\beta}^{ll'}(|\mathbf{r}^l - \mathbf{r}^{l'}|)\right|$, and the scalar value of $\left|\boldsymbol{\beta}^{ll'}\right|$ will depend only from the distance of the neighbor (displaced) sites.

---

† We should note that system (13), (14) in Fourier representation can be solved analytically for three symmetrical directions of wave vector $\mathbf{k}$. For $\mathbf{k}$ of general position, numerical calculation becomes necessary, so it is convenient to insert small inertia forces $\mu\ddot{P}^l$ and $\nu\ddot{Q}_{\alpha\beta}^l$ with $\mu$, $\nu \ll m$ into the left parts of (13) in order to use standard calculation programs. In the final analysis, we should reject characteristic oscillation with high frequencies of order $1/\sqrt{\mu}$ and $1/\sqrt{\nu}$.

So, the linear approximation with respect to displacements yields

$$\boldsymbol{\beta}^{ll'} = \frac{(\mathbf{r}^l - \mathbf{r}^{l'})}{r_0}\left|\boldsymbol{\beta}^{ll'}(r_0)\right| + \frac{\mathbf{u}^l - \mathbf{u}^{l'}}{r_0}\left|\boldsymbol{\beta}^{ll'}(r_0)\right| + \\ + \frac{(\mathbf{r}^l - \mathbf{r}^{l'})}{r_0^2}(\mathbf{u}^l - \mathbf{u}^{l'})(\mathbf{r}^l - \mathbf{r}^{l'})\frac{d}{dr}\left[\frac{\beta^{ll'}(r)}{r}\right]_{r_0}. \quad (15)$$

The first term vanishes after the summation over $l`$ due to central symmetry of the environment. Analogous principle can be applied to the term $D_{\alpha\beta}^l, Q_{\alpha\beta}^l$. According to (8), $D_{\alpha\beta}^l = \sum_{l'}^{n.n.} D_{\alpha\beta}^{ll'}$. Every $D_{\alpha\beta}^{ll'}$ term is determined only by atom positions $l'$ and $l''$. In coordinates $\xi\ \eta\ \zeta$, where $\xi$ axis is aligned with $\mathbf{r}^{l'} + \mathbf{u}^{l'} - \mathbf{r}^l - \mathbf{u}^l$, axial symmetry is an evident reason of the fact that mixed components of this tensor are equal to zero. $\sum_{\alpha=1}^{3.} \hat{Q}_{\alpha\alpha}^l = 0$, because it is a feature of $\hat{Q}_{\alpha\beta}^l$ operators being parts of definition of $D_{\alpha\beta}^l$ (8). That is why only one independent component exists

$$D_{\xi\xi}^{ll'} = -2D_{\eta\eta}^{ll'} = -2D_{\zeta\zeta}^{ll'}$$

The invariant $\sum_{\alpha\beta} D_{\alpha\beta}^{ll'} \cdot Q_{\alpha\beta}^l$ is written down first in these local coordinates. Then we should expand $D_{\xi\xi}^{ll'}$ up to terms linear with respect to displacements $\mathbf{u}^l - \mathbf{u}^{l'}$,

$$D_{\xi\xi}^{ll'} = D_{\xi\xi}^{ll'}(r_0) + \frac{1}{r_0} \cdot \frac{dD_{\xi\xi}^{ll'}}{dr}(\mathbf{u}^l - \mathbf{u}^{l'}) \cdot \mathbf{r}^{ll'}. \quad (16)$$

and transform the expression to the common crystal coordinates. The result is a bilinear form of $Q_{\alpha\beta}^l(u_\gamma^l - u_\gamma^{l'})$. Evidently,

$$\frac{1}{2}\sum_{l'} D_{\alpha\beta}^{ll'} \cdot Q_{\alpha\beta}^l = \frac{1}{2}\sum_{l'} D_{\xi\xi}^{ll'}\left[Q_{\xi\xi}^l - \frac{1}{2}(Q_{\eta\eta}^l + Q_{\zeta\zeta}^l)\right] = \\ = \frac{3}{4}a\sum_{l'} D_{\xi\xi}^{ll'} \cdot q_{\xi\xi}^l = \\ = \frac{3}{4}a\sum_{l'}\left[D_{\xi\xi}^{ll'}(r_0) + \frac{1}{ae\sqrt{2}} \cdot \frac{dD_{\xi\xi}^{ll'}}{dr}\bigg|_{r_0} \cdot (\mathbf{p}^l - \mathbf{p}^{l'})\mathbf{r}^{ll'}\right] \cdot q_{\xi\xi}^l. \quad (17)$$

Here all the variables have the same dimension with dipole moments $q_{\xi\xi}^l = \frac{1}{\grave{a}}Q_{\xi\xi}^l$; $\mathbf{p}^l = e\mathbf{u}^l$; $a$ is a half of a cubic cell edge. Now we transform a tensor component

$$q_{\xi\xi}^l = \sum_{\alpha\beta} q_{\alpha\beta}^l \cdot cos(\xi\alpha) \cdot cos(\xi\beta). \quad (18)$$

Substituting (18) to (17), we can take $cos(\xi\alpha) \cdot cos(\xi\beta)$ in the first approximation as $r_\alpha^{l'l} \cdot r_\beta^{l'l}/r_0^2$ in the term containing small value of $(\mathbf{u}^{l'} - \mathbf{u}^l) \cdot \mathbf{r}^{l'l}$, and the term of zero order $D_{\xi\xi}^{ll'}(r_0)$ requires them in the first order with respect to displacements

$$cos(\xi\alpha) = \frac{(r_\alpha^{l'l} + u_\alpha^{l'} - u_\alpha^l)}{|\mathbf{r}^{l'l} + \mathbf{u}^{l'} - \mathbf{u}^l|} = \\ = \frac{r_\alpha^{l'l}}{r_0} + \frac{u_\alpha^{l'} - u_\alpha^l}{r_0} - \frac{(\mathbf{u}^{l'} - \mathbf{u}^l) \cdot \mathbf{r}^{l'l} \cdot r_\alpha^{l'l}}{r_0^3}.$$

Thus,

$$cos(\xi\alpha)cos(\xi\beta) = \frac{r_\alpha^{l'l} \cdot r_\beta^{l'l}}{r_0^2} + \\ + \frac{(u_\alpha^{l'} - u_\alpha^l) \cdot r_\beta^{l'l} + (u_\beta^{l'} - u_\beta^l)r_\alpha^{l'l}}{r_0^2} - 2\frac{(\mathbf{u}^{l'} - \mathbf{u}^l) \cdot \mathbf{r}^{l'l} r_\alpha^{l'l} r_\beta^{l'l}}{r_0^4}. \quad (19)$$

As a result, the linear term with respect to quadruples in (7) has the form

$$\frac{1}{2}\sum_{\alpha\beta} D^{ll'}_{\alpha\beta} \cdot Q^l_{\alpha\beta} \to \frac{3a}{8e\sqrt{2}} \cdot \left.\frac{dD^{ll'}_{\xi\xi}(r)}{dr}\right|_{r_0} \sum_{\alpha\beta}^{(9)} q^l_{\alpha\beta} \sum_{i=1}^{6} e_{i\alpha} e_{i\beta} \cdot$$
$$\cdot (\mathbf{p}^{l+e_i} - \mathbf{p}^{l-e_i}) \cdot \mathbf{e}_i +$$  (20)
$$+ \frac{3}{8e} D_{\xi\xi}(r_0) \sum_{\alpha\beta}^{(9)} \sum_{i}^{6} q^l_{\alpha\beta} \begin{bmatrix} (\mathbf{p}_\alpha^{l+e_i} - \mathbf{p}_\alpha^{l-e_i}) e_{i\beta} + \\ + (\mathbf{p}_\beta^{l+e_i} - \mathbf{p}_\beta^{l-e_i}) e_{i\alpha} - \\ - e_{i\alpha} e_{i\beta} (\mathbf{p}^{l+e_i} - \mathbf{p}^{l-e_i}) \cdot \mathbf{e}_i \end{bmatrix}.$$

Here $\mathbf{e}_i$ are dimensionless vectors directed from $\mathbf{r}^l$ to the 6 nearest neighbors $\mathbf{r}^{l'}$

$$\mathbf{e}_1 = \mathbf{i} + \mathbf{j}; \quad \mathbf{e}_2 = \mathbf{i} - \mathbf{j}; \quad \mathbf{e}_3 = \mathbf{i} + \mathbf{\kappa};$$
$$\mathbf{e}_4 = \mathbf{i} - \mathbf{\kappa}; \quad \mathbf{e}_5 = \mathbf{j} + \mathbf{\kappa}; \quad \mathbf{e}_6 = \mathbf{j} - \mathbf{\kappa};$$

The other 6 neighbors determine $\mathbf{e}_i$ vectors and are explicitly taken into account in (20).

So, the theory will contain 4 parameters:

$$\left|\boldsymbol{\beta}^{ll'}\right|; \quad \frac{d}{dr}\left[\frac{1}{r_0}\boldsymbol{\beta}^{ll'}(r_0)\right]_{r_0}; \quad D^{ll'}_{\xi\xi}(r_0); \quad \frac{1}{r_0}\left[\frac{dD^{ll'}_{\xi\xi}(r_0)}{dr}\right]_{r_0}.$$

Finally, the expansion of $U_{sr}$ with respect to degrees of displacements in quadratic approximation (linear terms are zeroized because of the symmetry) will contain only squares of longitudinal $\mathbf{u}_{\text{II}}$ and transversal $\mathbf{u}_\perp$ (against $(\mathbf{r}^l - \mathbf{r}^{l'})$ line) differences displacements $\mathbf{u}^l - \mathbf{u}^{l'}$. The mixed term $\mathbf{u}_{\text{II}} \cdot \mathbf{u}_\perp$ vanishes because the transversal force $\frac{\partial U}{\partial \mathbf{u}_\perp}$ is absent when $\mathbf{u}_\perp = 0$.

That is why
$$U_{sr}(\mathbf{r}^l + \mathbf{u}^l - \mathbf{r}^{l'} - \mathbf{u}^{l'}) =$$
$$= const + \frac{f}{4}(\mathbf{u}^l - \mathbf{u}^{l'})^2 + \frac{d}{4}\frac{\left[(\mathbf{u}^l - \mathbf{u}^{l'})\mathbf{r}^{ll'}\right]^2}{r_0^2}, \quad (21)$$

where $f$ and $d$ are theory parameters. In the case of central short-range forces, they are expressed in terms of the first and the second derivative of $U_{sr}(\mathbf{r})$

$$f = \frac{1}{r_0}\cdot\left.\frac{dU_{sr}}{dr}\right|_{r_0}; \quad d = \left.\frac{d^2 U_{sr}}{dr^2}\right|_{r_0} - \frac{1}{r_0}\cdot\left.\frac{dU_{sr}}{dr}\right|_{r_0}. \quad (22)$$

Then the adiabatic potential has the form of a quadratic form of $\mathbf{p}^l$, $\mathbf{P}^l$, $q^l_{\alpha\beta}$, divided by the cubed length and we get

$$U = \frac{1}{a^3} \sum_l \left\{ \begin{array}{l} \dfrac{(\mathbf{P}^l)^2}{2A} + \sum_{l'}^{(12)} \left[ \begin{array}{l} \dfrac{h}{4}(\mathbf{p}^l - \mathbf{p}^{l'})\mathbf{P}^l + \dfrac{1}{2r_0^2} g \mathbf{P}^l \cdot \mathbf{r}^{ll'} (\mathbf{p}^l - \mathbf{p}^{l'}) \cdot \mathbf{r}^{ll'} + \\ \dfrac{H}{16}(\mathbf{p}^l - \mathbf{p}^{l'})^2 + \dfrac{G}{8r_0^2} \left\langle (\mathbf{p}^l - \mathbf{p}^{l'}) \cdot \mathbf{r}^{ll'} \right\rangle^2 \end{array} \right] + \\[2ex] + \dfrac{1}{2b}\left[ \sum_\alpha (q^l_{\alpha\alpha})^2 + \sum_{\alpha \neq \beta}^{(6)} (q^l_{\alpha\beta})^2 \right] + \dfrac{1}{4}\sum_l \sum_{i=1}^{6} \sum_{\alpha\beta}^{(9)} g^l_{\alpha\beta} \left[ \begin{array}{l} w e_{i\alpha} \cdot e_{i\beta}(\mathbf{p}^{l+e_i} - \mathbf{p}^{l'-e_i})\mathbf{e}_i + \\ + v(p_\alpha^{l+e_i} - p_\alpha^{l-e_i}) \cdot e_{i\beta} \end{array} \right] + \\[2ex] + \dfrac{1}{2} \sum_{l'} \left\langle \begin{array}{l} \dfrac{(\mathbf{p}^l - \mathbf{p}^{l'})^2}{2} \left[ B\left(\dfrac{a}{|\mathbf{r}^{ll'}|}\right)^8 + R\left(\dfrac{a}{|\mathbf{r}^{ll'}|}\right)^{10} + S\left(\dfrac{a}{|\mathbf{r}^{ll'}|}\right)^{12} \right] - \\ - \dfrac{\left[(\mathbf{p}^l - \mathbf{p}^{l'}) \cdot \mathbf{r}^{ll'}\right]^2}{r_0^2} \times \\ \times \left[ 4B\left(\dfrac{a}{|\mathbf{r}^{ll'}|}\right)^8 + 5R\left(\dfrac{a}{|\mathbf{r}^{ll'}|}\right)^{10} + 6S\left(\dfrac{a}{|\mathbf{r}^{ll'}|}\right)^{12} \right] \end{array} \right\rangle + \\[2ex] + \dfrac{1}{2}\sum_{l'} \left[ \begin{array}{l} \dfrac{\mathbf{P}^l \cdot \mathbf{P}^{l'}}{|\mathbf{r}^{ll'}|^3} - 3\dfrac{(\mathbf{P}^l \cdot \mathbf{r}^{ll'})(\mathbf{P}^l \cdot \mathbf{r}^{ll'})}{|\mathbf{r}^{ll'}|^5} - 2a \sum_{\alpha\beta} \dfrac{q^l_{\alpha\beta} \cdot X^{ll'}_\alpha \cdot P^{l'}_\beta}{|\mathbf{r}^{ll'}|^5} + \\ 5a \sum_{\alpha\beta\gamma} \dfrac{q^l_{\alpha\beta} \cdot X^{ll'}_\alpha \cdot X^{ll'}_\beta \cdot \mathbf{P}^l \cdot \mathbf{r}^{ll'}}{|\mathbf{r}^{ll'}|^7} + \\ + \dfrac{25}{6} a^2 \sum_{\alpha\beta} \dfrac{q^l_{\alpha\beta} \cdot q^{l'}_{\alpha\beta}}{|\mathbf{r}^{ll'}|^5} - \dfrac{5}{3} a^2 \sum_{\alpha\beta\gamma} \dfrac{q^l_{\alpha\beta} \cdot q^{l'}_{\beta\gamma} \cdot X^{ll'}_\alpha \cdot X^{ll'}_\gamma}{|\mathbf{r}^{ll'}|^7} + \\ + \dfrac{35}{6} a^2 \sum_{\alpha\beta\gamma\delta} \dfrac{q^l_{\alpha\beta} \cdot q^{l'}_{\gamma\delta} \cdot X^{ll'}_\alpha X^{ll'}_\beta X^{ll'}_\gamma X^{ll'}_\delta}{|\mathbf{r}^{ll'}|^9} \end{array} \right] \end{array} \right\}. \quad (23)$$

Here all the parameters are dimensionless

$$G = \frac{2d \cdot a^3}{e^2}; \quad H = \frac{4f \cdot a^3}{e^2}; \quad B = \frac{6C}{a^5 e^2};$$

$$R = \frac{8C'}{a^7 e^2}; \quad S = \frac{10C''}{a^9 e^2};$$

$$h = \frac{2\sqrt{2} \cdot \beta(r_0) \cdot a^2}{e}; \quad g = \frac{2a^3}{e}\left[\frac{d\beta(r)}{dr}\right]_{r_0} - \frac{h}{2};$$

$$A = \frac{\alpha}{a^3}; \quad b = \frac{2\beta_{44}}{a^5}; \quad (24)$$

$$w = \frac{3}{2e}\left[\frac{a}{\sqrt{2}}\frac{dD^{ll'}_{\xi\xi}(r)}{dr}\bigg|_{r_0} - D^{ll'}_{\xi\xi}(r_0)\right]; \quad v = \frac{3}{e} D^{ll'}_{\xi\xi}(r_0).$$

Using (20), we shall write equation of motion (13), (14) and look for the solution in the form of plane waves.

When differentiating with respect to $q^l_{\alpha\alpha}$, we take into account the relation $\sum_\alpha q^l_{\alpha\alpha} = 0$ and insert Lagrangian coefficients, Fourier-components labeled as λ. Because of the symmetry $q^l_{\alpha\beta} = q^l_{\beta\alpha}$, the result of differentiating with respect to $q^l_{\alpha\beta}$ at $\alpha \neq \beta$ should be halved. We suggest

$$p^l_\alpha = p_\alpha \cdot e^{-i\omega t + i\mathbf{K} \cdot \mathbf{r}^l}; \quad P^l_\alpha = P_\alpha \cdot e^{-i\omega t + i\mathbf{K} \cdot \mathbf{r}^l};$$
$$q^l_{\alpha\beta} = q_{\alpha\beta} \cdot e^{-i\omega t + i\mathbf{K} \cdot \mathbf{r}^l}. \quad (25)$$

Thus amplitudes $ð_\alpha, Ð_\alpha, q_{\alpha\beta}$ are determined by the system

$$\Omega^2 p_\alpha = hP_\alpha \mu(\mathbf{k}) +$$
$$+ g\left[ P_\alpha \cdot \nu_\alpha(\mathbf{k}) + \sum_{\alpha \neq \beta} P_\beta \cdot \tau_{\alpha\beta}(\mathbf{k}) \right] +$$
$$+ Hp_\alpha \mu(\mathbf{k}) + G\left[ p_\alpha \nu_\alpha(\mathbf{k}) + \sum_{\alpha \neq \beta} p_\beta \tau_{\alpha\beta}(\mathbf{k}) \right] +$$
$$+ \sum_\beta \left[ \chi^{(6)}_{\alpha\beta}(\mathbf{k}) \cdot B + \chi^{(8)}_{\alpha\beta}(\mathbf{k}) \cdot R + \chi^{(10)}_{\alpha\beta}(\mathbf{k}) \cdot S \right] p_\beta -$$
$$- i\left\{ \mathrm{v}\sum_\beta^{(3)} q_{\alpha\beta} \sigma_{\beta\beta}(\mathbf{k}) + w\left[ \sum_\beta^{(3)} \sigma_{\alpha\beta}(\mathbf{k}) \cdot q_{\beta\beta} + 2\sum_\beta^{(2)} q_{\alpha\beta} \sigma_{\beta\alpha}(\mathbf{k}) \right] \right\}$$

$$0 = \frac{\partial_\alpha}{A} + hp_\alpha \cdot \mu(\mathbf{k}) + gp_\alpha \cdot \nu_\alpha(\mathbf{k}) +$$
$$+ g\sum_\beta^{(2)} p_\beta \cdot \tau_{\alpha\beta}(\mathbf{k}) - \sum_\beta \varphi_{\alpha\beta}(\mathbf{k}) \cdot \partial_\beta -$$
$$- \sum_{\beta\gamma} \eta^{\alpha\beta\gamma}(\mathbf{k}) \cdot q_{\beta\gamma};$$

$$0 = \frac{1}{b} q_{\alpha\alpha} + i(w+\mathrm{v}) \eth_\alpha \sigma_{\alpha\alpha}(\mathbf{k}) +$$
$$+ iw\sum_\beta^{(2)} \eth_\beta \sigma_{\beta\alpha}(\mathbf{k}) + \sum_\gamma \eta^{\alpha\alpha\gamma}(\mathbf{k}) \cdot \partial_\gamma - \qquad (26)$$
$$- \sum_{\beta\gamma} \varsigma^{\alpha\alpha\beta\gamma}(\mathbf{k}) \cdot q_{\beta\gamma} + \lambda;$$

$$0 = \frac{1}{b} q_{\alpha\beta} +$$
$$+ i\left( p_\alpha \sigma_{\beta\alpha}(\mathbf{k}) + p_\beta \sigma_{\alpha\beta}(\mathbf{k}) \right) w +$$
$$+ \frac{i\mathrm{v}}{2} \left( p_\alpha \sigma_{\beta\beta}(\mathbf{k}) + p_\beta \sigma_{\alpha\alpha}(\mathbf{k}) \right) +$$
$$+ \sum_\gamma \eta^{\alpha\beta\gamma}(\mathbf{k}) \cdot P_\gamma - \sum_{\gamma\delta} \varsigma^{\alpha\beta\gamma\delta}(\mathbf{k}) \cdot q_{\gamma\delta}.$$

Here dimensionless frequencies $\Omega = \omega\sqrt{\dfrac{ma^3}{\mathring{a}^2}}$ and the following functions of dimensionless wave vector $\mathbf{k} = a\mathbf{K}$ are inserted.

$$\mu(\mathbf{k}) = 3 - \sum_{\gamma<\beta} \cos k_\gamma \cdot \cos k_\beta;$$
$$\nu_\alpha(\mathbf{k}) = 2 - \cos k_\alpha \sum_{\beta \neq \alpha} \cos k_\beta;$$
$$\tau_{\alpha\beta}(\mathbf{k}) = \sin k_\alpha \cdot \sin k_\beta; \qquad (27)$$
$$\sigma_{\alpha\beta}(\mathbf{k}) = \sin k_\alpha \cdot \cos k_\beta;$$
$$\sigma_{\alpha\alpha}(\mathbf{k}) = \sin k_\alpha \sum_{\alpha \neq \beta} \cos k_\beta.$$

They arise at the nearest neighbor summation. Comparatively long-order Van der Waals forces yield after summation the functions $\chi^{(6)}_{\alpha\beta}(\mathbf{k})$, $\chi^{(8)}_{\alpha\beta}(\mathbf{k})$, $\chi^{(10)}_{\alpha\beta}(\mathbf{k})$ determined by

$$\chi^{(n)}_{\alpha\beta}(\mathbf{k}) = \frac{1}{n}\left\{ \frac{\partial^2 F_n(\mathbf{k},\boldsymbol{\rho}) \cdot e^{i\mathbf{k}\boldsymbol{\rho}}}{\partial \rho_\alpha \partial \rho_\beta} - \frac{\partial^2 F_n(0,\boldsymbol{\rho})}{\partial \rho_\alpha \partial \rho_\beta} \right\}_{\boldsymbol{\rho}=0} \quad (28)$$

where $F_n(\mathbf{k},\boldsymbol{\rho}) = \sum_l{}' \dfrac{e^{i\mathbf{k}(\mathbf{l}-\boldsymbol{\rho})}}{|\mathbf{l}-\boldsymbol{\rho}|^n}; \quad n = 6, 8, 10.$

Finally, summarized Coulomb forces of long order after addition give the functions $\varphi_{\alpha\beta}(\mathbf{k})$, $\eta^{\alpha\beta\gamma}(\mathbf{k})$, $\varsigma^{\alpha\beta\gamma\delta}(\mathbf{k})$, that are the second, the third and the fourth derivatives of the function

$$S(\mathbf{k},\boldsymbol{\rho}) = \sum_l{}' \frac{e^{i\mathbf{k}(\mathbf{l}+\boldsymbol{\rho})}}{|\mathbf{l}+\boldsymbol{\rho}|};$$

with multipliers 1, 1/3! и 1/(3!)2, correspondingly. The functions $\varphi_{\alpha\beta}(\mathbf{k})$ were derived in the source paper by Tolpygo K.B. [11], ad their values for 28 points of 1/48 of Brilloin zone calculated with using Evald's method are presented in [19]. $\chi^{(6)}_{\alpha\beta}(\mathbf{k})$ functions are derived in [29] by transformation of $F_6(\mathbf{k},\boldsymbol{\rho})$ sums according to Embersleben's formulae [30]. Analogous calculation of $\chi^{(2)}_{\alpha\beta}(\mathbf{k})$ and $\chi^{(10)}_{\alpha\beta}(\mathbf{k})$, as well as their values for symmetrical directions of $\mathbf{k}$ are presented in [24]. $\eta^{\alpha\beta\gamma}(\mathbf{k})$ functions for 8 points in $\mathbf{k}$ - space (for [100] and [111] directions) can be found in [31], and $\varsigma^{\alpha\beta\gamma\delta}(\mathbf{k})$ for three directions are contained in [32]. The system (26) is a totality of 12 equations for three $p\alpha$ components, three $P\alpha$ and six $q_{\alpha\beta}$. The condition $\sum_l q_{\alpha\alpha} = 0$ allows us to exclude an additional variable $\lambda$.

## II. Elastic properties and Cauchy relation of rare-gas crystals under high pressure. Many-body interaction and quadruple deformation of electron shells.

### 1. Investigation of long-wave lattice oscillations.

Now we shall consider (26) in approximation of $k \langle\langle 1$, so we expand all the functions (27), (28) with respect to $k$ degrees up to a term $\Box\, k^2$ inclusive. Thus, we get

$$\mu(\mathbf{k}) = k^2,\quad \nu_\alpha(\mathbf{k}) = \frac{k^2}{2} + \frac{k_\alpha^2}{2},$$
$$\sigma_{\alpha\alpha}(\mathbf{k}) = 2k_\alpha,\quad \sigma_{\alpha\beta}(\mathbf{k}) = k_\alpha, \qquad (29)$$
$$\tau_{\alpha\beta}(\mathbf{k}) = k_\alpha \cdot k_\beta.$$

$$\varphi_{\alpha\beta}(\mathbf{k}) = \frac{2\pi}{3}\delta_{\alpha\beta} - 2\pi \cdot \frac{k_\alpha \cdot k_\beta}{k^2} - 0.2371 k^2 \cdot \delta_{\alpha\beta} + 0.28999 k_\alpha \cdot k_\beta + 0.42128 k_\alpha^2 \cdot \delta_{\alpha\beta} \qquad (30)$$

$$i\eta^{\alpha\alpha\alpha}(\mathbf{k}) = k_\alpha(\frac{\pi}{3} \cdot \frac{k_\alpha^2}{k^2} - 0.41484),$$
$$i\eta^{\alpha\alpha\beta}(\mathbf{k}) = k_\beta(\frac{\pi}{3} \cdot \frac{k_\alpha^2}{k^2} - 0.31592),\ \alpha \neq \beta,$$
$$i\eta^{\alpha\beta\gamma}(\mathbf{k}) = \frac{\pi}{3} \cdot \frac{k_\alpha \cdot k_\beta \cdot k_\gamma}{k^2},\ \alpha \neq \beta \neq \gamma,$$
$$\zeta^{\alpha\alpha\alpha\alpha}(\mathbf{k}) = -0.1565 + \frac{\pi}{18} \cdot \frac{k_\alpha^4}{k^2},$$
$$\zeta^{\alpha\alpha\beta\beta}(\mathbf{k}) = 0.07824 + \frac{\pi}{18} \cdot \frac{k_\alpha^2 \cdot k_\beta^2}{k^2},\ \alpha \neq \beta.$$

Other combinations of signs
$$\zeta^{\alpha\beta\gamma\delta}(\mathbf{k}) = \frac{\pi}{18} \cdot \frac{k_\alpha \cdot k_\beta \cdot k_\gamma \cdot k_\delta}{k^2},$$
$$\alpha = \beta = \gamma \neq \delta,\ \alpha = \beta \neq \gamma \neq \delta.$$

Further
$$\chi_{\alpha\alpha}^{(6)}(\mathbf{k}) = -0.26247 \cdot k^2 - 0.71820 \cdot k_\alpha^2,$$
$$\chi_{\alpha\beta}^{(6)}(\mathbf{k}) = -1.12718 \cdot k_\alpha \cdot k_\beta,\ \alpha \neq \beta;$$
$$\chi_{\alpha\alpha}^{(8)}(\mathbf{k}) = -0.18951 \cdot k^2 - 0.36463 \cdot k_\alpha^2,$$
$$\chi_{\alpha\beta}^{(8)}(\mathbf{k}) = -0.64568 \cdot k_\alpha \cdot k_\beta,\ \alpha \neq \beta; \qquad (31)$$
$$\chi_{\alpha\alpha}^{(10)}(\mathbf{k}) = -0.12523 \cdot k^2 - 0.20133 \cdot k_\alpha^2,$$
$$\chi_{\alpha\beta}^{(10)}(\mathbf{k}) = -0.37870 \cdot k_\alpha \cdot k_\beta,\ \alpha \neq \beta.$$

Substituting (31) in (26), we see that $P_\alpha$ is of order of $k^2$ with respect to $p_\alpha$ and $q_{\alpha\beta}$ is about $k$. As $P_\alpha$ enters the first group of equations (26) being multiplied by $k^2$, they can be neglected in this approximation. Eliminating $q_{\alpha\alpha}$ and $q_{\alpha\beta}$ from the last group of equations of (26) and substituting them in the first one, we get equations similar to ones of elasticity theory.

$$p_\alpha \Omega^2 = p_\alpha k^2 \begin{bmatrix} H + \frac{1}{2}G - 0.26247B \\ -0.18951R - 0.125235S \\ -\frac{(w+v)^2}{1/b + 0.15649} \end{bmatrix} +$$
$$(\mathbf{pk})k_\lambda \begin{bmatrix} G - 1.12718B - \\ 0.64568R - 0.37870S - \\ -\frac{(w+v)^2}{1/b+0.15649} + \frac{1}{3}\frac{(w+2v)^2}{1/b+0.23474} \end{bmatrix} + \quad (32)$$
$$p_\alpha k_\alpha^2 \begin{bmatrix} -\frac{1}{2}G + 0.40898B + \\ +0.28085R + 0.17737S + \\ +\frac{2(w+v)^2}{1/b+0.15649} - \frac{(w+2v)^2}{1/b+0.23474} \end{bmatrix}.$$

If we compare them with the equations of macroscopic theory of elasticity at zero pressure

$$\rho p_\alpha \omega^2 = B_{44} p_\alpha k^2 + (B_{12} + B_{44})(\mathbf{pk})k_\alpha + \\ + (B_{11} - B_{12} - 2B_{44}) p_\alpha k_\alpha^2 \qquad (33)$$

and insert the dimensional values of $\omega$ and $\mathbf{k}$ to (32), we obtain the following expressions for Birch elastic moduli $B_{ik}$ ‡ associated with the modules Bragger $C_{ik}$

---

‡ While deriving (32) and (34), we suggest the pressure equal to zero, too. So, a definite condition is imposed to the parameters of the theory. It will be formulated and used below (see.(36)) for the case of central forces.

following formula
$B_{\alpha\beta\gamma\vartheta} = C_{\alpha\beta\gamma\vartheta} - p(\delta_{\alpha\gamma}\delta_{\beta\vartheta} + \delta_{\alpha\vartheta}\delta_{\beta\gamma} - \delta_{\alpha\beta}\delta_{\gamma\vartheta})$:

$$B_{44} = \frac{e^2}{2a^4}\left[\begin{array}{l}H + \frac{1}{2}G - 0.26247B - \\ -0.18951R - 0.125235S - \\ -\frac{(w+v)^2}{1/b + 0.15649}\end{array}\right];$$

$$B_{12} = \frac{e^2}{2a^4}\left[\begin{array}{l}\frac{1}{2}G - H - 0.86471B - \\ -0.45617R - 0.25347S + \\ +\frac{1}{3}\frac{(w+2v)^2}{1/b - 0.23474}\end{array}\right];$$

$$B_{11} = \frac{e^2}{2a^4}\left[\begin{array}{l}G + H - 0.98067B - \\ -0.55434R - 0.32656S - \\ -\frac{2}{3}\frac{(w+2v)^2}{1/b - 0.23474}\end{array}\right].$$
(34)

In central force approximation, $H$ can be expressed through the rest of parameters if the energy of the lattice has a minimum at the experimental value of the lattice constant. Uniform compression or stretching of the crystal zeroized all $P_\alpha$, $q_{\alpha\beta}$ and the energy per a cell is

$$U = 6U_{sr}(a\sqrt{2}) - \frac{1}{2a^6}\left[\begin{array}{l}F_6(0.0)\cdot C + \\ +\frac{1}{a^2}F_8(0.0)\cdot C' + \frac{1}{a^4}F_{10}(0.0)\cdot C''\end{array}\right].$$ (35)

where $F_6(0.0)$, $F_8(0.0)$, $F_{10}(0.0)$ according to (28) are lattice sums of $1/r^6$, $1/r^8$, $1/r^{10}$, being equal to 1,80674; 0,80001; 0,38472, correspondingly.

Differentiating $U$ with respect to $a$ and equating to zero, we get $f$ (22) and $d$ (24) by definition, so

$$H = -\frac{1}{6}[B\cdot F_6(0.0) + R\cdot F_8(0.0) + S\cdot F_{10}(0.0)]. (36)$$

Inserting (36) to (34) and simplifying the formula, we get

$$B_{44} = \frac{e^2}{2a^4}\left[\begin{array}{l}\frac{1}{2}G - 0.56359B - 0.32284R - \\ -0.18935S - \frac{(w+v)^2}{1/b + 0.15649}\end{array}\right];$$

$$B_{12} = B_{44} + \frac{e^2}{2a^4}\left[\begin{array}{l}\frac{(w+v)^2}{1/b + 0.15649} + \\ +\frac{1}{3}\cdot\frac{(w+2v)^2}{1/b - 0.23474}\end{array}\right];$$

$$B_{11} = \frac{e^2}{2a^4}\left[\begin{array}{l}G - 1.2818B - 0.68768R - \\ -0.39074S - \frac{2}{3}\cdot\frac{(w+2v)^2}{1/b - 0.23474}\end{array}\right]. (37)$$

It is seen that Cauchy relation $C_{12} = C_{44}$ ($B_{12} = B_{44} + 2p$) at central short-range forces can be obtained only at neglect of quadruple atom deformation. If the last fact is essential, as seen from (37) $B_{12} > B_{44}$. The inequality is valid for heavy RGC at $p=0$. For Ar and Ne, a number of authors mentioned different values of elastic moduli, one data gave at p=0 $B_{12} > B_{44}$, other data supposed $B_{12} \langle B_{44}$. We think that the inverse inequality is connected with noncentral $U_{sr}$ ((36) becomes invalid and many-body forces are present). This problem is considered in details in [23, 32] and in the chapter below.

## 2. Deviation from Cauchy relation in heavy RGC

A series of our papers [33, 34, 35, 36] contained the theory of elastic properties of RGC under pressure based on non-empirical calculation of short-range potential of repulsion $U_{sr}$ including both the pair one (accounting for the first and the second neighbors), and the many-body one. In [35, 36], short-range many-body forces caused by overlap of atomic electron shells are studied within the frameworks of K.B. Tolpygo's model with no account taken of the deformation of electron shells. Regard for many-body interaction in harmonic approximation changes two-body interaction making it a non-central one and determines the presence of many-body terms in crystal oscillation equations. Many-body forces arising because of orthogonalization of wave functions change the behavior of dispersion curves at whole range of $k$ and violate Cauchy relation, in particular. A good agreement of theoretical and experimental deviation from Cauchy relation was obtained for Ar in a wide pressure range.

For heavy RGC, it is necessary to take into account quadruple deformation of electron shells. The deviation from Cauchy relation written with the help of Birch moduli will take the form:

$$\delta = B_{12} - B_{44} - 2p =$$
$$= \frac{e^2}{2a^4}\left[2\delta H - V_t + \frac{1}{2}T + \frac{1}{3}V_q - 4R_t\right]. \quad (38)$$

The parameters of many-body interaction were obtained in [35]. They are expressed in terms of overlap integral $S = S_{zz}^{ll}$ and its derivatives $S_i$, $f_i$, so there is an opportunity to calculate these parameters for every crystal of Ne-Xe series individually:

$$\delta H = -64a^3 S(r_1)\begin{bmatrix} 2S_2(r_1)f(r_2) + \\ +3S(r_1)f_2(r_2) - 2S_1(r_1)f_1(r_2) \end{bmatrix}, (39)$$

$$\delta G = -64a^3 \begin{bmatrix} 2S(r_1)S_3(r_1)f(r_2) + \\ +S_1^2(r_1)f(r_2) + \\ +4S(r_1)S_1(r_1)f_1(r_2) + \\ +9S^2(r_1)f_3(r_2) \end{bmatrix}, \quad (40)$$

where $r_1 = a\sqrt{2}$ is the distance between the nearest neighbors and $r_2 = a\sqrt{6}/2$, $f = S(r_2)/r_2$.

$$V_t = 128\frac{a^3}{e^2}\left[S(r)\frac{a}{r}\frac{dS(r)}{dr}\right]_{r=a\sqrt{2}} \left[\frac{a}{R}\frac{df\left(\frac{R}{2}\right)}{dR}\right]_{R=a\sqrt{6}}, (41)$$

$$R_t = -\frac{a^2}{6e^2}\frac{dU_t(a)}{da} > 0. \quad (42)$$

The parameters $T$ and $V$ are expressed in terms of the parameters of quadruple deformation $w$ and $v$ (24).

Our calculations [34] of elastic properties of Xe were carried out using pair non-empirical potential $V_{sr}$ calculated accurately within $S^2$ in approximation of the nearest and second-nearest neighbors (see [33] and [34] in details). However, we had no experiment to compare with our results at the moment. Fig.1 presents the experiment dated by 2009 [5] and our calculations of Birch moduli $B_{ij}$ [34]. As seen, the agreement is good and the due regard to the second-nearest neighbors makes an essential contribution, so it is necessary under high pressure.

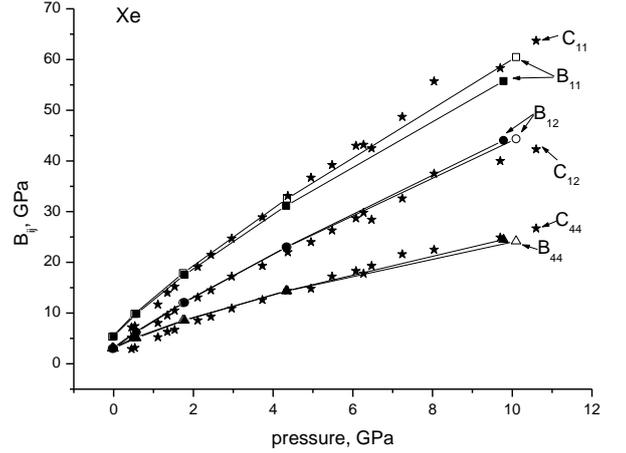

Fig. 1. Pressure dependence of the Birch modules $B_{ij}$ for Xe. Legend: squares is our $B_{11}$ calculation in the M3 model accounting for the nearest neighbors (solid symbols) and the second-nearest neighbors (open symbols) [34], the circles and the triangles are the same for $B_{12}$ and $B_{44}$, respectively, and the stars present the experiment [5].

Giving a good description of the equation of state and the elastic moduli, the theory based on the pair potential yields $C_{12} = C_{44}$ and it can not depict valuable deviation from CR observed at the experiment for all the crystals irrespective of the type of the chemical bond at zero and non-zero pressure.

Here we shall calculate the parameters of many-body interaction for Xe analogously to [35] using (39), (40), (41), (42) and evaluate the parameters of quadruple interaction. As seen from (8), compression dependence of $D_{\alpha\beta}^{ll'}$ can be obtained after the calculation of the matrix element $\langle i0|H_{sr}^{ll'}|00\rangle$. To evaluate, we suggest [18]

$$\langle i0|H_{sr}^{ll'}|00\rangle \approx \langle 00|H_{sr}^{ll'}|00\rangle = V_{sr}^{ll'} \approx A\frac{S^2}{R}, \quad (43)$$

where $R$ is the distance between the atoms $l$ and $l'$ (for the nearest neighbors $R = r_1$), $A$ is a constant. Besides, we suppose $T \approx 8V$ [23, 24].

Fig.2 demonstrates the compression dependence of the required parameter $V$, $u = \Delta V/V_0$ ($\Delta V = V_0 - V(p)$, $V_0$ - is the volume at $p = 0$) at varied $A$. It is seen that the best result is obtained at $A = 0.8$. The table demonstrates many-body and quadruples parameters as well as the contribution to $\delta$ due to many-body interaction $\delta_t$ and quadruple interaction $\delta_q$ and $\delta_{exp}$ [5].

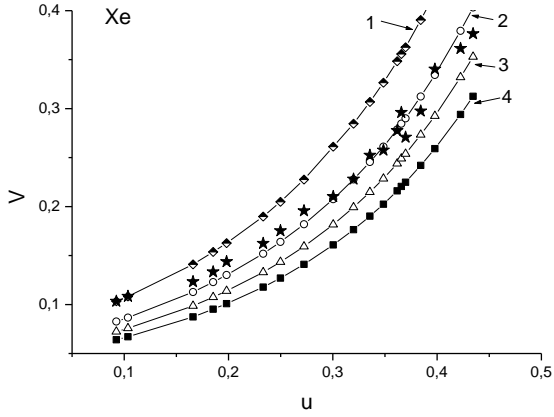

Fig. 2. Pressure dependence of the quadrupole parameter $V$ for the Xe. Legend: the curves 1, 2, 3, and 4 are the coefficients A = 1, 0.8, 0.75 and 0.62 (see formula (43)), respectively. The stars are calculation of $V$ by (38) at $\delta = \delta_{exp}$ [5].

As seen in Fig.3 [7, 6], the deviation from CR for Xe caused only by many-body interaction does not conform with experimental $\delta_{exp}$ at all, as opposed to Ar [35]. Our calculation of $\delta_{theor} = \delta_t + \delta_q$ accounting for quadruple deformation $\delta_q$ depicts CR deviation in a good agreement with the experiment. *Ab initio* calculations of density functional theory (DFT) [6] agree with the experiment only near $p = 0$. With pressure increase, the discrepancy becomes more essential. The same tendency can be traced in results of [7] where the calculations were based on empirical potentials in many-body model.

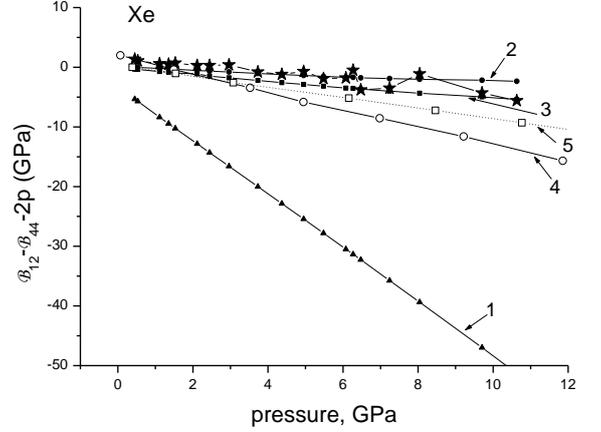

Fig. 3. Pressure dependence of deviation from Cauchy relation $\delta$ (38) for Xe. Legend: 1 is our calculation of $\delta_t$ taking into account the three-body interaction only ($V = T = 0$), 2 and 3 are our calculation of $\delta_{theor} = \delta_t + \delta_q$ taking into account the quadrupole interaction $V$ (see 2 and 3 on the Fig. 2), 4 is the calculation in the DFT [6], 5 is the calculation of the many-body model with empirical potentials [7], the stars are the experiment [5].

**Conclusion**

The series of papers "Elementary oscillations in rare-gas crystals". 1-3» [37, 38, 39] considered non-adiabatic effects, i.e. electron-phonon interactions determined by electron shell deformation in dipole approximation. This way corresponds to the account for the lowest terms with respect to non-adiabaticity. As known [40], they do not contribute to elasticity moduli. The next order, i.e. the consideration of electron-phonon interaction determined by electron shell deformation in quadruple approximation results in appearance of the corresponding terms in expressions for elasticity moduli (34). They make a lower contribution in comparison with the pair potential but they are comparable with the contribution of many-body interaction (the parameters |V0| and V are of the same order). It is especially seen at the analysis of CR deviation of $\delta_{exp}$ for heavy RGC, in any case. We should note that *ab initio* calculations in density functional theory do not reproduce $\delta_{exp}$ for Kr and Xe [29].

In this manner, we have shown that quadruple interaction plays an important role in elastic properties of RGC under high pressure and it should be taken into account side by side with many-body interaction.

The present *ab initio* investigation of CR violation gave us an opportunity to establish the nature and balance of forces forming crystal properties under high pressures. Thus it was shown that CR violation in RGC is determined by two factors:

(i) three-body forces generated by overlapping of electron shells of "rigid" atoms in the crystal;

(ii) electron-phonon interaction manifesting itself as electron shell deformation of a quadruple-type atom at nucleus displacement.

We should note in conclusion that construction of a dynamical theory within the model of «rigid» atoms is incorrect even at $p = 0$, although in this case electron-phonon interaction is small and, as a consequence, deformation of atomic electron shells is small. Low energy of interatomic interaction of closed spherically symmetric shells results in the fact that atoms weakly deform each other. But this effect does not give grounds to ignore these deformations because only this parameter is responsible for the bond of atoms in a crystal as seen by the example of Van der Waals forces.

**Table.** Dimensionless parameters of many-body ($\delta G$, $\delta H$, $V_t$, $R_t$) and quadruple ($V$) interaction and deviation from Cauchy relation [GPa] depending on the pressure P [GPa] (the compression $u=\Delta V/V_0$)

| P | u | R | $\delta G$ | $\delta H$ | $V_0$ | $R_t$ | $\delta_t$ | V(0.8) | $\delta_q$ | $\delta_{theory}$ | $\delta_{exp}$ |
|---|---|---|---|---|---|---|---|---|---|---|---|
| 0,451 | 0,0924 | 7,9257 | 0,351978 | -0,18425 | -0,09746 | 0,021697 | -5,3359 | 0,08267 | 5,342111 | 0,006211 | 1,34 |
| 0,53 | 0,1036 | 7,8929 | 0,368636 | -0,19354 | -0,10207 | 0,028668 | -5,70777 | 0,086608 | 5,6899701 | -0,0178 | 1,06 |
| 1,111 | 0,166 | 7,7054 | 0,478205 | -0,25385 | -0,13232 | 0,030766 | -8,37128 | 0,112764 | 8,156493 | -0,21479 | 0,6 |
| 1,351 | 0,1853 | 7,6455 | 0,518521 | -0,27851 | -0,14345 | 0,033766 | -9,44837 | 0,122934 | 9,174136 | 0,27423 | 0,5 |
| 1,531 | 0,1982 | 7,6049 | 0,549733 | -0,2954 | -0,15156 | 0,035959 | -10,2574 | 0,130135 | 9,920388 | -0,337 | 0,7 |
| 2,112 | 0,2332 | 7,4926 | 0,634534 | -0,34614 | -0,17547 | 0,042666 | -12,8354 | 0,1519 | 12,28958 | -0,5452 | 0,3 |
| 2,442 | 0,2498 | 7,4382 | 0,680616 | -0,3735 | -0,18824 | 0,046314 | -14,3024 | 0,163909 | 13,65388 | -0,64854 | 0,3 |
| 2,961 | 0,2724 | 7,3627 | 0,747580 | -0,4138 | -0,20665 | 0,051789 | -16,5819 | 0,181975 | 15,78984 | -0,79206 | 0,4 |
| 3,732 | 0,3004 | 7,267 | 0,843454 | -0,4716 | -0,23309 | 0,059627 | -20,0151 | 0,207575 | 18,97857 | -1,03653 | -0,8 |
| 4,369 | 0,3199 | 7,1988 | 0,910179 | -0,51627 | -0,2531 | 0,065795 | -22,8438 | 0,227663 | 21,6148 | -1,22901 | -1,2 |
| 4,951 | 0,3356 | 7,143 | 0,982001 | -0,55438 | -0,26966 | 0,071423 | -25,4235 | 0,245487 | 24,0439 | -1,37964 | -0,7 |
| 5,481 | 0,3485 | 7,0965 | 1,033971 | -0,58935 | -0,28498 | 0,076093 | -27,7973 | 0,261226 | 26,26358 | -1,53371 | -1,9 |
| 6,078 | 0,3617 | 7,0482 | 1,092629 | -0,62732 | -0,30175 | 0,081394 | -30,483 | 0,2787 | 28,79562 | -1,68741 | -1,8 |
| 6,27 | 0,3657 | 7,0335 | 1,111130 | -0,63911 | -0,30682 | 0,083074 | -31,3462 | 0,284622 | 29,65505 | -1,69114 | -0,5 |
| 6,473 | 0,3698 | 7,0183 | 1,130659 | -0,65151 | -0,31215 | 0,08484 | -32,262 | 0,290099 | 30,48809 | -1,77389 | -3,8 |
| 7,242 | 0,3843 | 6,964 | 1,201931 | -0,69779 | -0,33203 | 0,091438 | -35,7572 | 0,312364 | 33,86298 | -1,89418 | -3,5 |
| 8,041 | 0,3979 | 6,9124 | 1,271700 | -0,74307 | -0,36096 | 0,098027 | -39,3637 | 0,334342 | 37,34121 | -2,02252 | -1,1 |
| 9,704 | 0,4225 | 6,8169 | 1,407829 | -0,83361 | -0,38841 | 0,111301 | -46,9759 | 0,379418 | 44,79926 | -2,17663 | -4,3 |
| 10,63 | 0,4345 | 6,7694 | 1,475367 | -0,88275 | -0,40846 | 0,118571 | -51,31609 | 0,403276 | 48,96827 | -2,34782 | -5,6 |

Note: $\delta_t$ is the deviation from Cauchy relation by the rate of many-body interaction, $\delta_q$ is the deviation from CR determined by electron shell deformation in quadruple approximation, $\delta_{theory}= \delta_q + \delta_t$, $\delta_{exp}$ is the experimental deviation from CR [5]

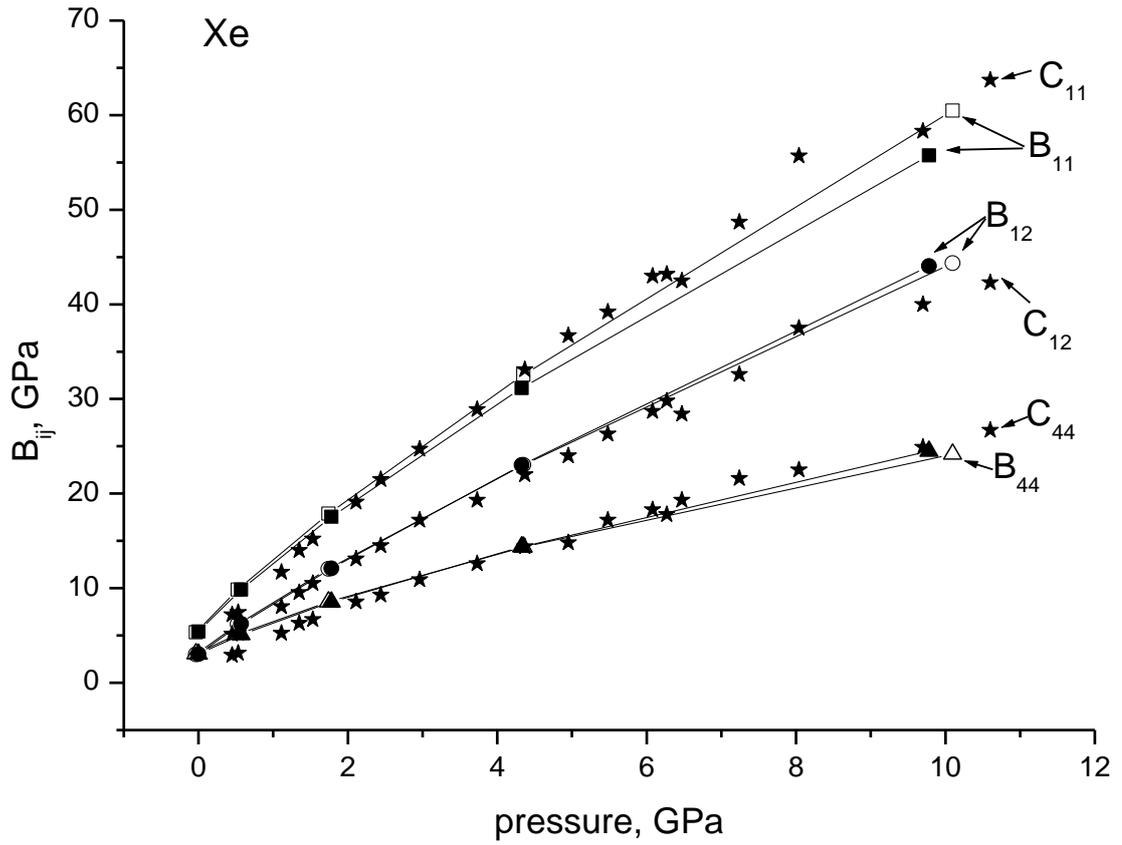

Fig. 1. Pressure dependence of the Birch modules $B_{ij}$ for Xe. Legend: squares is our $B_{11}$ calculation in the M3 model accounting for the nearest neighbors (solid symbols) and the second-nearest neighbors (open symbols) [34], the circles and the triangles are the same for $B_{12}$ and $B_{44}$, respectively, and the stars present the experiment [5].

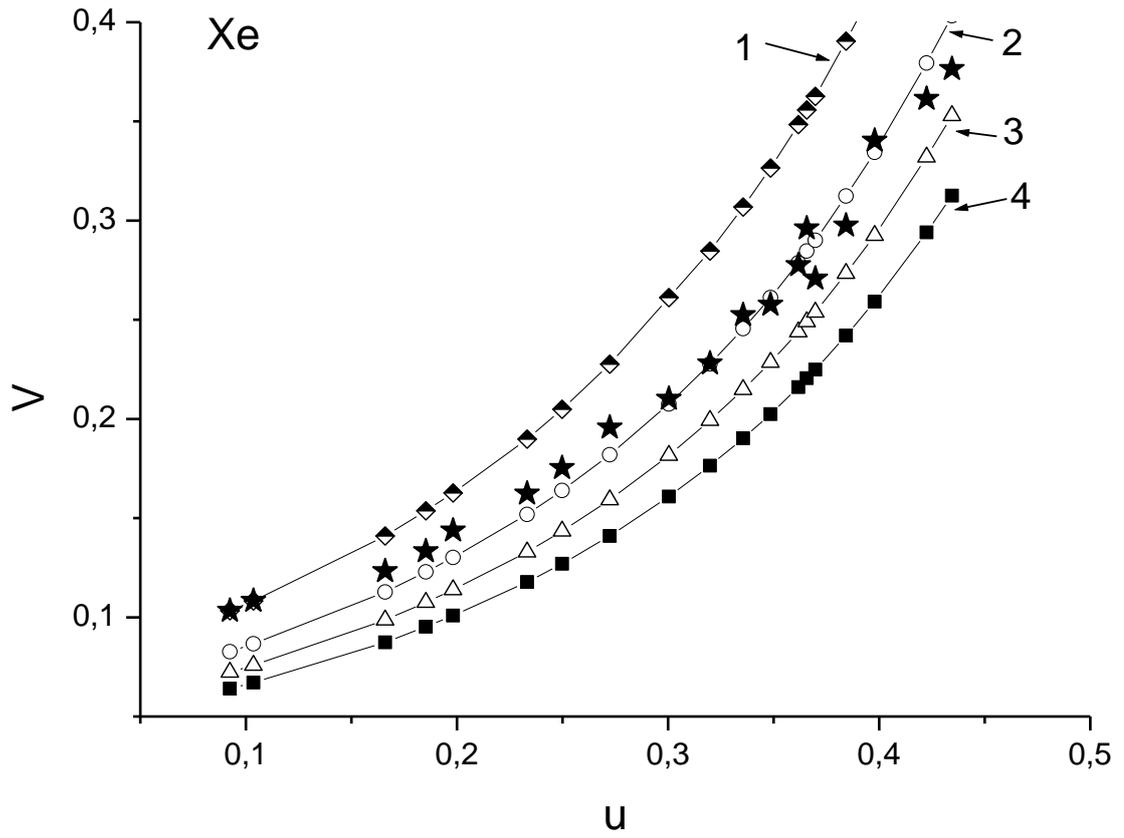

Fig. 2. Pressure dependence of the quadrupole parameter *V* for the Xe. Legend: the curves 1, 2, 3, and 4 are the coefficients A = 1, 0.8, 0.75 and 0.62 (see formula (43)), respectively. The stars are calculation of *V* by (38) at $\delta = \delta_{exp}$ [5].

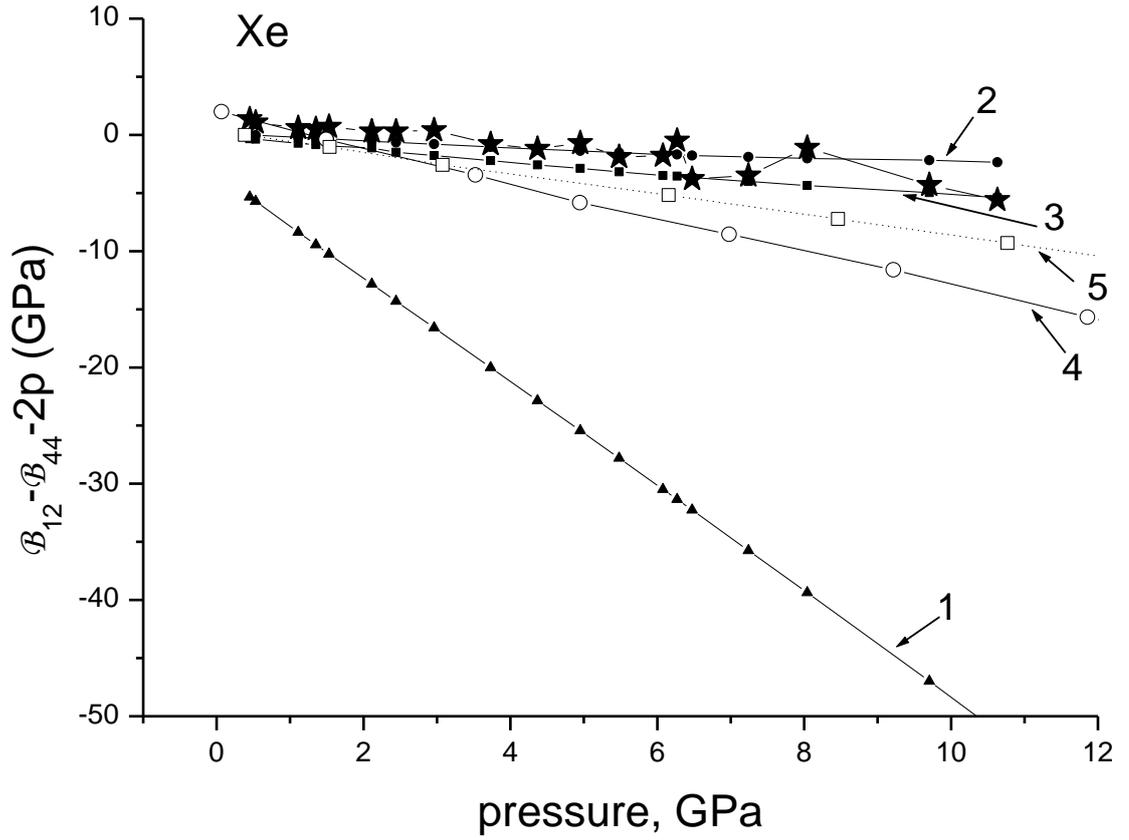

Fig. 3. Pressure dependence of deviation from the Cauchy relation $\delta$ (38) for Xe. Legend: 1 is our calculation of $\delta_t$ taking into account the three-body interaction only ($V = T = 0$), 2 and 3 are our calculation of $\delta_{theor} = \delta_t + \delta_q$ taking into account the quadrupole interaction $V$ (see 2 and 3 on the Fig. 2), 4 is the calculation in the DFT [6], 5 is the calculation of the many-body model with empirical potentials [7], the stars are the experiment [5].